\documentclass[prc,amsmath,twocolumn,showkeys,superscriptaddress,preprintnumbers,floatfix]{revtex4}
\usepackage{gensymb}
\usepackage{graphicx,color}
\usepackage{amssymb}
\usepackage{enumerate}
\usepackage{verbatim}
\usepackage{natbib}
\usepackage{dsfont}
\usepackage{float}
\usepackage{subcaption}
\usepackage[normalem]{ulem}
\usepackage[outdir=./]{epstopdf}
\usepackage{caption}
\usepackage{subcaption}
\usepackage{textcomp}
\usepackage{gensymb}
\usepackage{amsbsy}
\usepackage{mathtools}
\newcommand{\p}{{p}}

\newcommand{\bv}{{\mathbf{b}}}

\newcommand{\y}{{\mathbf{y}}}

\newcommand{\model}{{\pmb{\sigma}}}
\newcommand{\thetav}{{\pmb{\rho}}}
\newcommand{\onetheta}{{\rho}}
\newcommand{\gammav}{{\pmb{\gamma}}}
\newcommand{\Cv}{{\mathbf{C}}}

\newcommand{\Wv}{{\mathbf{W}}}
\newcommand{\Kv}{{\mathbf{K}}}
\newcommand{\Bv}{{\mathbf{B}}}
\newcommand{\cv}{{\mathbf{c}}}
\newcommand{\rhov}{{\pmb{\zeta}}}
\newcommand{\muv}{{\pmb{\mu}}}
\newcommand{\wv}{{\mathbf{w}}}
\newcommand{\epsilonv}{{\pmb{\epsilon}}}
\newcommand{\pcaepsilonv}{{\pmb{e}}}
\newcommand{\realR}{{\mathbb{R}}}
\newcommand{\bone}{{\mathbf{1}}}
\newcommand{\Sigmav}{{\pmb{\Sigma}}}

\newcommand{\angular}{\frac{{\rm d} \sigma}{{\rm d}\Omega}} 
\newcommand{\energy}{\frac{{\rm d} \sigma}{\rm d E}}

\begin{document}
  \newcommand {\nc} {\newcommand}
  \nc {\Sec} [1] {Sec.~\ref{#1}}
  \nc {\IBL} [1] {\textcolor{black}{#1}} 
  \nc {\IR} [1] {\textcolor{red}{#1}} 
  \nc {\IB} [1] {\textcolor{blue}{#1}} 
  \nc {\IG} [1] {\textcolor{green}{#1}}
  \nc {\beq} {\begin{eqnarray}}
  \nc {\eeq} {\nonumber \end{eqnarray}}
  \nc {\eeqn}[1] {\label {#1} \end{eqnarray}}
  \nc {\ve} [1] {\mbox{\boldmath $#1$}}

\title{Uncertainty Quantification in Breakup Reactions}

\author{{\"O}.~Sürer}
\email{ozgesurer2019@u.northwestern.edu}
\affiliation{Northwestern-Argonne Institute of Science and Engineering (NAISE), Northwestern University, Evanston, IL 60208, USA}

\author{F.M.~Nunes}
\email{nunes@frib.msu.edu}
\affiliation{Facility for Rare Isotope Beams, Michigan State University, East Lansing, MI~48824, USA}
\affiliation{Department of Physics and Astronomy, Michigan State University, East Lansing, MI~48824, USA}

\author{M. Plumlee}
\email{mplumlee@northwestern.edu}
\affiliation{Industrial Engineering and Management Sciences Department, Northwestern University, Evanston, IL 60208, USA}
\affiliation{Northwestern-Argonne Institute of Science and Engineering (NAISE), Northwestern University, Evanston, IL 60208, USA}

\author{S.M.~Wild}
\email{wild@anl.gov}
\affiliation{Mathematics and Computer Science Division, Argonne National Laboratory, Lemont, IL 60439, USA}
\affiliation{Northwestern-Argonne Institute of Science and Engineering (NAISE), Northwestern University, Evanston, IL 60208, USA}

\date{\today}

\begin{abstract} 
Breakup reactions are one of the favored probes to study loosely bound nuclei, particularly those in the limit of stability forming a halo. In order to interpret such breakup experiments, the continuum discretized coupled channel method is typically used. In this study, the first Bayesian analysis of a breakup reaction model is performed. We use a combination of statistical methods together with a three-body reaction model (the continuum discretized coupled channel method) to quantify the uncertainties on the breakup observables due to the parameters in the effective potential describing the loosely bound projectile of interest. The combination of tools we develop opens the path for a Bayesian analysis of not only breakup processes, but also a wide array of complex processes that require computationally intensive reaction models. 
\end{abstract}

\maketitle

\section{Introduction} \label{sec:intro}

The study of rare isotopes has unveiled an array of surprising exotic phenomena, arising from the subtle interplay of the different forces acting on loosely bound nucleons. In some instances, these nucleons are allowed to reside at unusually large distances from the center of mass of the system, creating a halo. In these so-called halo nuclei \cite{hove2018}, there is a strong decoupling of the core degrees of freedom and the degrees of freedom associated with the halo nucleons.  
Halo nuclei can be found for many different isotopic chains, both at the neutron-rich limit as well as the proton-rich extreme. For example, in the Carbon isotopic chain, one finds the exotic $^{9}$C \cite{fukui2012} in the proton-rich side and the two-neutron halo $^{22}$C \cite{tanaka2010,pinilla2016,nagahisa2018} on the neutron dripline; in the Boron isotopic chain, while $^8$B exhibits a well studied proton halo (e.g.,  \cite{rangel2016,mazzocco2019,wang2021}), $^{19}$B has a neutron halo \cite{cook2020,casal2020}.

Halo nuclei are primarily studied through reactions and, over recent decades, a variety of experimental programs at many rare isotope facilities worldwide have unveiled their properties (e.g., \cite{davids2001,nakamura2009,chen2016,heine2017,cook2020}). Breakup reactions offer a unique probe into these systems: since the halo nucleus breaks up so easily, the cross sections for the process are large. From the analysis of breakup experiments we learn about the halo properties of the ground state (particularly the asymptotic normalization coefficient (ANC)) as well as the low-lying continuum including resonances (e.g., \cite{capel2006}). From breakup measurements, we can also infer capture reactions for astrophysics \cite{nunes2020}.

Reaction theory is needed in order to connect the breakup measurement to the quantities of interest, be it specific properties of the projectile or a capture reaction of astrophysical interest. One challenge faced when modeling breakup is that, for most cases, it  becomes computationally intensive to evaluate observables given theory parameters. 
It has been shown that the simple perturbative approaches are unreliable \cite{capel2012} and instead there are large coupled equations that need to be solved, requiring a significant amount of run time (e.g., \cite{deltuva2009,upadhyay2012,hlophe2019}). 
As a consequence, the current predictions include no uncertainty quantification.

Progress on uncertainty quantification in reaction theory has offered important insights, particularly concerning the effective potential, a strong contributor to the theoretical error \cite{lovell2017,lovell2018}. Many of the recent uncertainty quantification studies involve a Bayesian analysis of elastic scattering within the optical model, and at times, the propagation of uncertainties on the optical potential to specific reaction channels that can be modeled in a simple perturbative description \cite{king2019,catacora2019,lovell2020,catacora2021,whitehead2021}. It is important to go beyond the first-order reaction models covered in these pioneering works, and enable a Bayesian analysis of the state-of-the-art reaction theory models \cite{deltuva2009,upadhyay2012,hlophe2019}. Obviously, from the computational cost, it is not feasible to do statistical calculations (such as Markov chain Monte Carlo) directly on the model. 
Emulators that can efficiently leverage model outputs are called for to solve this problem.

Recently, emulators for two-body elastic scattering have been developed based on the eigenvector continuation method \cite{furnstahl2020,drischler2021,melendez2021}. To describe the breakup dynamics one needs at least a three-body theory, as is the case for the Continuum Discretized Coupled Channel method (CDCC) \cite{cdcc}, the working horse in most analyses of breakup reactions.
Despite some preliminary work \cite{zhang2021}, it is not straightforward to extend such an eigenvector continuation method to the general three-body problem. Instead, a standard emulation by Gaussian Processes (GPs) (see \cite{santner2018design, gramacy2020surrogates}) should be considered. GP emulators often outperform other statistical learning tools for these types of tasks \citep{myren2021comparison}.

In this work, we implement an emulator based on GPs for breakup cross sections and couple this to a Bayesian framework. We focus on the breakup reaction of $^8$B$ + ^{208}$Pb$ \rightarrow ^7$Be$ + p + ^{208}$Pb at $80$ MeVA, which has been studied extensively  \cite{davids2001,capel2007}. The emulator is trained with CDCC calculations and the sharpness and accuracy of the uncertainty quantification of the emulator is studied. We then use the emulator for calibration, and extract posterior distributions for the parameters of the effective interaction between the halo nucleon and the core. Lastly, we obtain credible intervals for relevant reaction observables.

\section{Model description} \label{sec:model}

We now detail our reaction model and Bayesian calibration. 

\subsection{Modeling breakup reactions}

Breakup reactions of halo nuclei of the type $a+t \rightarrow c+v+t$ are typically formulated in terms of a three-body problem $c+v+t$ with the projectile assumed to have two-body structure $a=c+v$. The three-body Hamiltonian
\beq
H_{3B}=T_r + T_R + V_{cv}(\ve r)+U_{ct}(\ve{R}_{ct})+U_{vt}(\ve{R}_{vt})
\eeqn{eq:h3b}
is written in terms of the respective pairwise interactions, the interaction describing the internal states of the projectile $V_{cv}$ and the optical potentials describing the scattering of the core (valence nucleons) and the target $U_{ct}$ ($U_{vt}$). Here, $T_r$ and $T_R$ are the kinetic energy operators associated with the Jacobi coordinates: $\ve r$ is the relative coordinate between the core and the valence nucleon and  $\ve R$ is the relative coordinate connecting the center of mass of the projectile and the target. 

The CDCC method \cite{cdcc} expands the full three-body wavefunction $\Psi(\ve r, \ve R)$ in terms of the eigenstates of the projectile $\varphi_i(\ve r)$ defined by
\beq
\left[ T_r +  V_{cv}(\ve r) \right] \varphi_{i}(\varepsilon,\ve{r})=\varepsilon\varphi_{i}(\varepsilon,\ve{r}).
\eeqn{eq:hproj}
For halo nuclei, typically there is one bound state $i=0$ and the remaining states lie in the continuum. Because the continuum solutions of Eq.~\ref{eq:hproj} are not square integrable, the CDCC method discretizes the continuum into energy bins and averages the scattering states over the energy or momentum as described in \cite{book}. When this expansion is introduced in the three-body Sch\"odinger equation $H_{3B} \Psi = E \Psi$, one obtains the so-called CDCC equations, which consist of tightly coupled second-order differential equations that must be solved with the appropriate boundary conditions.
For more details, please consult \cite{book,nunes99}.

There are three pairwise interactions in this model, introducing close to $30$ parameters. Clearly breakup data alone would not be sufficient to inform all these parameters. Since this is the first Bayesian study for this type of reaction, here we choose to focus only on the uncertainty from the core-valence effective interaction, and thus fix the optical potentials to those used in a previous study \cite{mortimer2002,capel2007}. The effective interaction $V_{cv}$ has been shown to have a very important effect in the breakup cross section distributions, renormalizing the cross section according to the ANC of the ground state \cite{capel2006} and thus we expect the breakup data will be highly informative for the $V_{cv}$ parameters.
This $V_{cv}$ interaction is typically parameterized by a Woods-Saxon form \cite{capel2007}.
We take the radius  and diffuseness for the spin-orbit term to be the same as that for the central term. We then vary the Coulomb radius, the Woods-Saxon radius, the diffuseness, and the depth of the spin-orbit term ($R_C, R_{ws}, a_{ws}, V_{so}$)  by adjusting the depth of the central interaction $V_{ws}$ to reproduce the binding energy of the $^8$B system with $\varepsilon = 0.137$ MeV.  For simplicity, the same interaction is used to produce both the $^{8}$B ground state and all the continuum states. 

For a combination of parameters ($R_C, R_{ws}, a_{ws}, V_{so}$) in the physical intervals defined in Table~\ref{tab:parameters}, the corresponding CDCC cross sections are obtained using {\sc frescox} \cite{fresco}. Given that the calculations of \cite{capel2007} reproduce well the data, the parameters used therein, and in particular the core-valence interaction in the projectile ($R_C=R_{ws}=2.391$ fm, $a_{ws}=0.52$ fm, and $V_{so}=4.898$ MeV), are used to produce our mock observable data, namely the breakup angular distributions and the breakup energy distributions. In the next subsection, we discuss the implementation of the emulators for these breakup calculations.
\begin{table}[tbh!]
    \centering
        \caption{Model parameters and their ranges. \label{tab:parameters}}
    \begin{tabular}{c|| c l} 
    Parameter & Label & Range $[\underline{\onetheta_i}, \overline{\onetheta_i}]$ \\[0.5ex] 
    \hline\hline
     $R_{C}$ & $\onetheta_1$ & [2, 3] (fm) \\ 
     $R_{ws}$ & $\onetheta_2$ & [2, 3] (fm) \\
     $a_{ws}$ & $\onetheta_3$ & [0.4, 0.9] (fm) \\
     $V_{so}$ & $\onetheta_4$ & [2, 8] (MeV)
 \\
    \end{tabular}
\end{table}

\subsection{Bayesian calibration of breakup reactions}
\label{sec:bayesian_calibration}

After reviewing Bayesian calibration principles and notation, we present our methodology for emulation of CDCC calculations.

\subsubsection{Bayesian calibration}

We represent a CDCC simulation with a mathematical function $\model(\cdot)$ that takes values of parameters $\thetav$ and returns output $\model(\thetav) \in \realR^d$.
We note that $\model(\thetav)$ represents multidimensional output observables, which include both cross section angular distributions and energy distributions. $\model(\thetav)$ should not be confused with total or reaction cross sections. We collect the four parameters in Table~\ref{tab:parameters} in the vector $\thetav = (\onetheta_1, \ldots, \onetheta_4)$. To emulate and then calibrate the breakup reactions, we use breakup angular $\angular$ and energy $\energy$ distributions. We denote the dimension of the $\angular$ and $\energy$ outputs by $d_\Omega$ and $d_E$, respectively, with $d = d_\Omega + d_E$. 

Next we describe the model space needed for convergence of the CDCC calculations. 
The model space is optimized per observable and we find that the bin discretization needed for producing converged $\angular$ is different than that needed for converged $\energy$.
For the angular distribution, the continuum is discretized up to a maximum relative energy of $\varepsilon_{\rm max} = 10$ MeV, and all spin/parity states are included up to a relative angular momentum between core and valence of $\ell=3$.
For each partial wave, the continuum is discretized into energy bins evenly spaced from 0 MeV to 3 MeV with a step size of 0.25 MeV, and from 3 MeV to 10 MeV with a step size of 1 MeV. 
The energy distributions also contain all partial waves up to $\ell=3$, but we take the maximum relative energy of $\varepsilon_{\rm max} = 3$ MeV with a uniform grid with a step size of $0.20$ MeV. This finer discretization is needed to capture the details of the peak of the distribution.

A calibration process uses observations from the real system (here, the mock data), denoted by vector $\y = (y_1, \ldots, y_d) = ( \{ \angular_{\rm mock} \}, \{ \energy_{\rm mock} \} )$ with $d=d_{\Omega} + d_E$, to constrain the uncertainty on the input parameters $\thetav$. 
This is done using a statistical model of the form
\begin{equation}
    \y = \model(\thetav) + \epsilonv, 
\end{equation}
where $\epsilonv\sim {\rm MVN}(\mathbf{0}, \Sigmav)$ denotes the residual error following a multivariate normal distribution (MVN) with mean $\mathbf{0}$ and covariance matrix $\Sigmav$.   

In the Bayesian framework, the model parameters are viewed as random variables, and the posterior probability density $\p(\thetav|\y)$ indicates the posterior probability of the parameters $\thetav$ given the observations $\y$. Based on Bayes' rule, the posterior density has the form 
\begin{equation}
    \p(\thetav|\y) = \frac{\p(\y|\thetav) \p(\thetav)}{\p(\y)} \propto \p(\y|\thetav) \p(\thetav), \label{eq:posterior}
\end{equation}
where $\p(\thetav)$ is the prior probability for the parameter $\thetav$ and $\p(\y|\thetav)$ is the likelihood function indicating how the model output of a set of parameters $\model(\thetav)$ agrees with the observations $\y$. Assuming $\epsilonv \sim {\rm MVN}(\mathbf{0}, \Sigmav)$, the likelihood satisfies
\begin{equation}
   \p(\y|\thetav) \propto |\Sigmav|^{-1/2} \exp\bigg(-\frac{1}{2} (\y - \model(\thetav))^\top \Sigmav^{-1} (\y - \model(\thetav )) \bigg). \label{eq:truelike}
\end{equation}

Markov chain Monte Carlo (MCMC) techniques \cite{Gelman2004} (e.g., Metropolis-Hastings algorithm) are generally used to sample from the posterior distribution in Eq.~\ref{eq:posterior}. During an MCMC procedure, an evaluation of the likelihood $\p(\y|\thetav)$ in Eq.~\ref{eq:truelike} is needed for each candidate parameter $\thetav$. The likelihood requires the simulation output $\model(\thetav)$ to be evaluated at a given parameter setting $\thetav$. MCMC procedures usually require thousands or millions of evaluations of the simulation depending upon the model's complexity and the number of input parameters considered. Therefore, a direct evaluation of a reaction model with MCMC techniques becomes computationally challenging when substantial computational time is required to produce a single model evaluation, as is the case for CDCC calculations (each run described here takes approximately 25 hours on a single core). 

Emulators are computationally efficient approximations of the input–output relationships for expensive simulations, and they have been used to address the computational challenges associated with executing the simulation run. To train an emulator, a sample of parameters is generated, and the corresponding simulation outputs are collected. Once an emulator is built, it can be used to efficiently predict the simulation outputs for parameter values that are not in the training set. In this study, we use GP emulators since, in addition to a predictive mean, they can produce a predictive variance, which allows us to quantify the uncertainty on emulation predictions. For an untried parameter $\thetav$, the GP provides a probabilistic representation of the simulation output with mean $\muv^{\rm GP}(\thetav)$ and covariance $\Sigmav^{\rm GP}(\thetav)$ based on training data from a simulation experiment.

Once the emulator is built,
for any parameter $\thetav$ 
one can 
apply Eqs.~\ref{eq:posterior}~and~\ref{eq:truelike} to
approximate the posterior $\p(\thetav|\y)$ via
\begin{align}
    p(\thetav|\y) \propto & |\mathbf{V}(\thetav)|^{-1/2}\exp\left(-\frac{1}{2} \mathbf{m}(\thetav)^\top \mathbf{V}(\thetav) ^{-1} \mathbf{m}(\thetav)\right) p(\thetav),
\label{eq:emu_posterior}
\end{align}
where, for our GP emulator, the quantities $\mathbf{V}(\thetav)$ and $\mathbf{m}(\thetav)$ are defined as
\begin{align}
\mathbf{V}(\thetav) = \Sigmav + \Sigmav^{\rm GP}(\thetav) \text{ and } \mathbf{m}(\thetav) = \y - \muv^{\rm GP}(\thetav).
\label{eq:v-m}
\end{align}
In this study, we employ an MCMC method to draw approximate posterior samples from Eq.~\ref{eq:emu_posterior}. Such sampling employs the cheap GP emulator in place of the expensive CDCC simulation, which would require many years of computation. The details of the sampler are explained in Sec.~\ref{sec:results}.

\subsubsection{Emulating breakup reactions}
\label{sec:emulators}

We generated three separate GP emulators for the breakup cross sections, which were then used for calibration. The first emulator considers only breakup angular distributions $\angular$. The second emulator takes into account only the breakup energy distributions $\energy$. The third emulator includes both the angular and energy distributions. For each of these emulators, we use the same set of uncertain parameters as inputs to the CDCC calculations. 
Principal component analysis (PCA) \cite{Ramsay97functionaldata} is used to project the high-dimensional outputs into a low-dimensional space where the projection is a collection of latent outputs. Then, each latent output is modeled using an independent GP model (see details in Apps.~\ref{appendix:emulator}--\ref{appendix:hyper}). 

\begin{figure}[t]
    \begin{center}
        \includegraphics[width=0.48\textwidth]{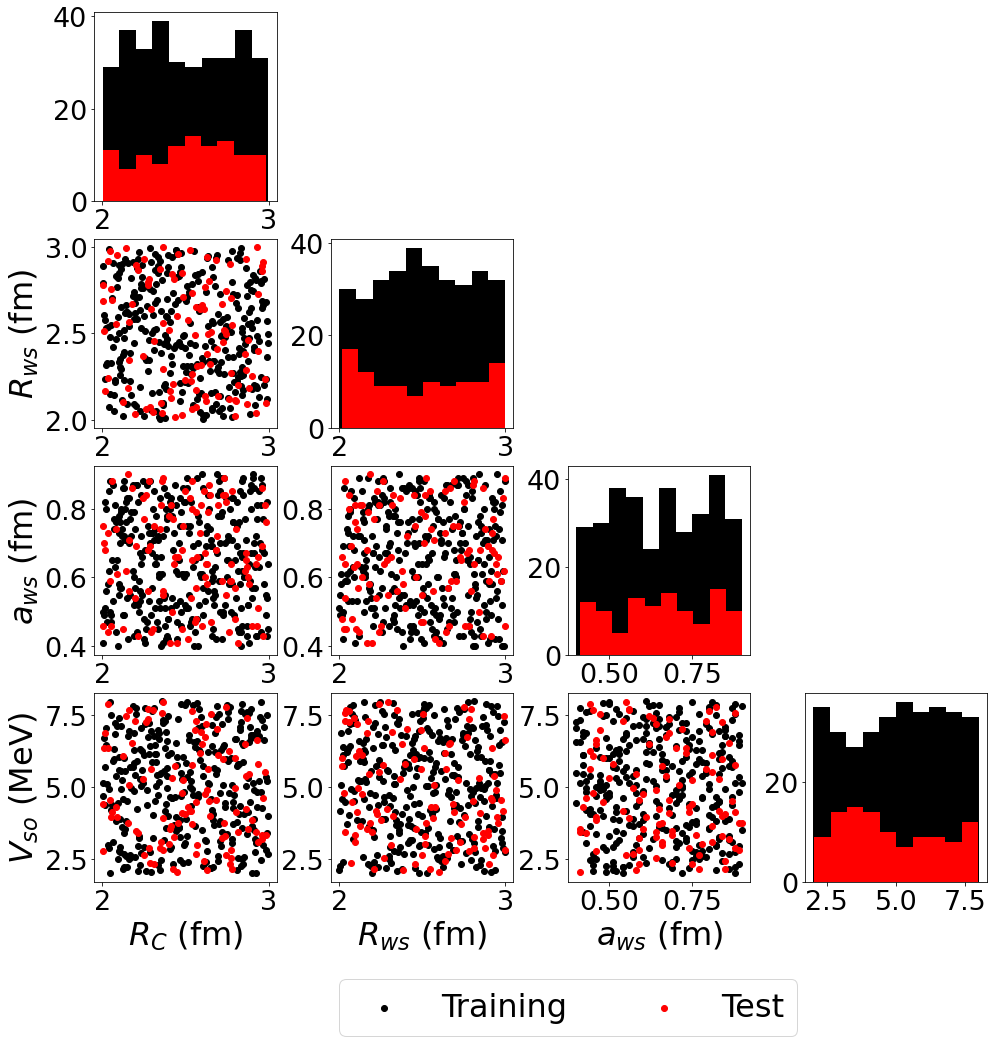}
    \end{center}
    \caption{Space-filling Latin hypercube sample (LHS) used to specify the set of breakup simulation runs to train the GP emulator (black) and to test the emulator (red). The 2-d projections of this design are shown for each pair of parameters.}
    \label{fig-sample}
\end{figure}

Since computing a single CDCC simulation via {\sc frescox} takes 25 hours on average, we use a GP-based emulator as a cheaper proxy to the CDCC simulation, as discussed before. We collect data via Latin hypercube sampling (LHS) (see the texts \cite{santner2018design, gramacy2020surrogates}) based on the ranges in Table~\ref{tab:parameters}. For evaluation purposes, we randomly split $75 \%$ of these data as training data and the remaining as test data; we then fit an emulator using the training data and evaluate the predictive quality of the trained emulator using the test data. Both the training and test data are illustrated in Fig.~\ref{fig-sample}; diagnostic plots are provided in Sec.~\ref{sec:predictions}. 

Once the quality of the emulator has been established, the final emulators are trained with a total of $n=500$ samples, collected via LHS. There were no additional difficulties found for the angular distribution emulator; however, for the energy distribution, we obtained outputs that differed tremendously from the mock data in some regions of the parameter space (see App.~\ref{appendix:pca} for more details). This is due to the existence of resonances in the core-valence system that greatly augment the sensitivity to parameters and reduce the precision in the resulting emulator. To tackle this problem, we developed a filter based on the deviation from the mock data of the predicted energy distributions: we discard from the training set any case in which $\energy$ is larger than the threshold of $1000$ mb/MeV (corresponding to roughly five times the peak of the ``real'' energy distribution).
Discarding parameter settings based on simulation outputs \cite{Surer2021} has been shown to provide emulators with better accuracy in the reduced space. As a consequence of filtering, the number of samples ultimately included in the training is $n=393$. For the third emulator, where the cross sections for angular and energy distributions are jointly considered, we also use $n = 393$ samples.
Simulation outputs before and after filtering are provided in App.~\ref{appendix:pca}.

\begin{figure}[ht]
    \begin{center}
        \includegraphics[width=0.44\textwidth]{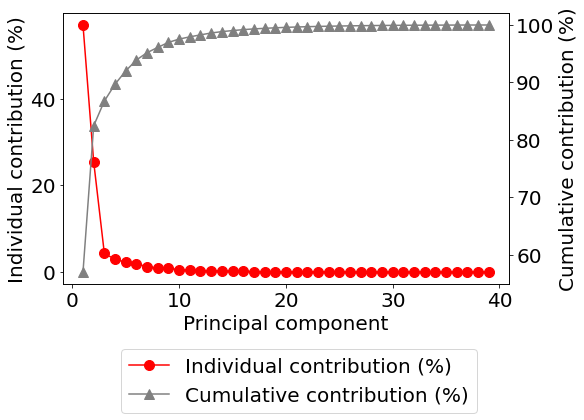}
    \end{center}
    \caption{The individual and cumulative contributions of principal components to the complete data set (angular distribution).}
    \label{fig-PCAexplained_angular}
\end{figure}

\begin{figure}[ht]
    \begin{center}
        \includegraphics[width=0.44\textwidth]{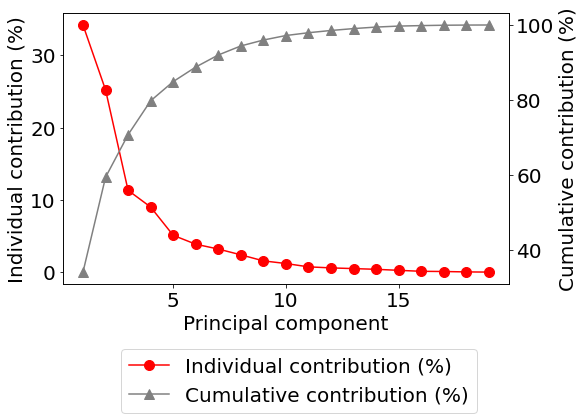}
    \end{center}
    \caption{The individual and cumulative contributions of principal components to the complete data set (energy distribution).}
    \label{fig-PCAexplained_energy}
\end{figure}

For the rest of this section, let $\{\thetav_1^{\rm tr}, \ldots, \thetav_n^{\rm tr}\}$ denote the ${n}$ parameter samples used to train each of the emulators. For each $\thetav^{\rm tr}_i$, $i=1, \ldots, n$, we obtain the simulation outputs (CDCC cross sections) via {\sc frescox} and use them as training data. The simulation outputs are represented in a $d \times n$ matrix $\Xi = [\model(\thetav_1^{\rm tr}), \ldots, \model(\thetav_{n}^{\rm tr})]$. We standardize each row of $\Xi$ so that it has zero mean and unit variance. After standardization, we employ PCA to reduce the dimension of the data from $d$ to $q \leq d$. We let $q_\Omega$, $q_E$, and $q_{\rm tot}$ denote the number of principal components used to fit an emulator for $\angular$, $\energy$, and both $\angular$ and $\energy$ as observables, respectively. PCA reduces the ($d_\Omega = 401$)- and ($d_E = 20$)-dimensional cross sections to $q_{\Omega} = 39$ and $q_{E} = 19$, respectively; this reduction is able to capture $99 \%$ of the variance. Note that for energy distribution there is no significant dimensionality reduction since all dimensions contribute to the variance of data. Figs.~\ref{fig-PCAexplained_angular}--\ref{fig-PCAexplained_energy} illustrate how much variance each principal component explains.

To combine $\angular$ and $\energy$ cross sections, we first reduce the dimension of $\angular$ (i.e., $d_\Omega = 401$) to 25 to treat $\angular$ and $\energy$ in a fair manner when computing the likelihood. To do that, we consider the cross sections up to the scattering angle of $3\degree$, which reduces the dimension from 401 to 50. This is a reasonable assumption because these breakup simulations are forward focused, and typically this is the angular range measured by detectors. We also find that including the larger angles does not modify the results. 
We then take $d_\Omega=25$ evenly spaced angles within these $3\degree$. The final set contains $d_{\rm tot} = 45$ (i.e., 25 for $\angular$ and 20 for $\energy$) dimensional observables. During emulation, PCA further reduces this set $d_{\rm tot} = 45$ to $q_{\rm tot} = 32$.

The $d \times n$ CDCC cross section matrix $\Xi$ is represented in the $q$-dimensional space via the $q \times n$ matrix $\Wv = \Bv^\top \Xi$ where the $d \times q$ matrix $\Bv = [\bv_1, \ldots, \bv_q]$ stores the orthogonal basis vectors $\{\bv_1, \ldots, \bv_q\}$. We obtain $\Wv$ with PCA as described in App.~\ref{appendix:emulator}, and fit a GP for each of $q$ latent outputs (i.e., each row of $\Wv$) such that
\begin{equation}
    w_i(\cdot) \sim \text{GP}(\gamma_i, C_i(\cdot, \cdot)),
    \label{eq:gp_prior}
\end{equation}
where $\gamma_i$ is the mean and $C_i(\cdot, \cdot)$ is the covariance function. The details of GP emulator are given in Apps.~\ref{appendix:emulator}--\ref{appendix:hyper}. The full implementation of this GP emulator for multidimensional simulation outputs is provided in the Python package \texttt{surmise} \cite{surmise2021}. 

\subsubsection{Predicting simulation outputs with emulators}
\label{sec:predictions}

As mentioned above, we fit three different emulators using training data. The next step is to test each of the emulators at any test parameter $\thetav$ to check whether they are able to predict the CDCC calculations. Let the $n$-dimensional vector $\wv_i$ be $\wv_i = (w_i(\thetav_1^{\rm tr}), \ldots, w_i(\thetav_{n}^{\rm tr}))$ for $i = 1, \ldots, q$, and let the $n \times n$ matrix $\Cv_{i}$ be the covariance matrix resulting from applying Eq.~\ref{eq:covariance} to each pair of the parameter settings $\thetav_1^{\rm tr}, \ldots, \thetav_{n}^{\rm tr}$. Following the results on normal distribution presented in \cite{Rasmussen2005}, one can derive that
\begin{equation}
    w_i(\thetav)|\wv_i \sim \text{N}(\nu_i(\thetav), \varsigma^2_i(\thetav,\thetav)), 
\end{equation}
with mean $\nu_i(\thetav)$ and variance $\varsigma^2_i(\thetav,\thetav)$ given by
\begin{equation}
    \begin{aligned}
        & \nu_i(\thetav) = \hat{\gamma}_i + \cv_i(\thetav)^\top \Cv_{i}^{-1} (\wv_i - \hat{\gamma}_{i} \bone_n) \\
        & \varsigma^2_i(\thetav,\thetav) = C_i(\thetav, \thetav) - \cv_i(\thetav)^\top \Cv_{i}^{-1} \cv_i(\thetav). \label{eq:meanvar_latent}
    \end{aligned}
\end{equation}
Here, the $n$-dimensional vector $\cv_i(\thetav)$ is obtained by evaluation of the covariance in Eq.~\ref{eq:covariance} between $\thetav$ and the training points $\{\thetav_1^{\rm tr}, \ldots, \thetav_n^{\rm tr}\}$ and $\hat{\gamma}_i$ is the estimated parameter for the mean function of a GP (see App.~\ref{appendix:hyper}).

For prediction at any test point $\thetav$, first, we obtain the mean $\nu_i(\thetav)$ and variance $\varsigma^2_i(\thetav,\thetav)$ in Eq.~\ref{eq:meanvar_latent} for each of the corresponding latent outputs for $i = 1, \ldots, q$. Then, these are transformed back to the original high-dimensional space through the inverse PCA transformation as follows. 
Define the $q$-dimensional vector $\muv(\thetav) = (\nu_1(\thetav), \ldots, \nu_q(\thetav))$ and $q \times q$ diagonal matrix $ \mathcal{S}(\thetav)$ with diagonal elements $\varsigma^2_i(\thetav,\thetav)$ for $i = 1, \ldots, q$. Then, due to the inverse PCA transformation, we obtain
\begin{equation}
    \model(\thetav) \sim \text{MVN}(\muv^{\rm GP}(\thetav), \Sigmav^{\rm GP}(\thetav))
    \label{emu_final}
\end{equation}
where $\muv^{\rm GP}(\thetav) \coloneqq \Bv \muv(\thetav)$ is the emulator predictive mean and $\Sigmav^{\rm GP}(\thetav) \coloneqq \Bv\mathcal{S}(\thetav)\Bv^\top$ is the covariance matrix. By plugging $\muv^{\rm GP}(\thetav)$ and $\Sigmav^{\rm GP}(\thetav)$ in Eq.~\ref{eq:v-m} we obtain $\mathbf{m}(\thetav)$ and $\mathbf{V}(\thetav)$. Then from Eq.~\ref{eq:emu_posterior}, we compute the posterior $p(\thetav|\y)$.
In this way, by employing the MCMC method on the emulators, we extract posterior distributions for the parameters of the effective interaction between the halo nucleon and the core without running the computationally expensive CDCC calculations when doing the MCMC sampling.

\begin{figure}
     \begin{subfigure}{0.48\columnwidth}
         \includegraphics[width=1\textwidth]{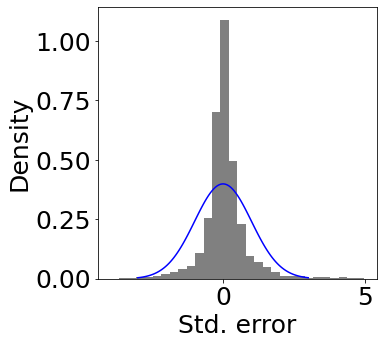}
         \caption{Standardized errors.} 
     \end{subfigure}
     \begin{subfigure}{0.48\columnwidth}
         \includegraphics[width=1\textwidth]{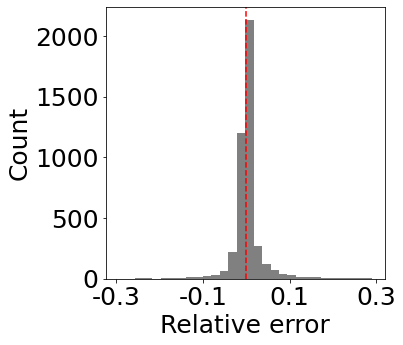}
         \caption{Relative errors.}
     \end{subfigure}
        \caption{Diagnostic plots to check the quality of the emulator using test parameters $\thetav_1^{\rm test}, \ldots, \thetav_{l}^{\rm test}$ and their corresponding simulation outputs $\model(\thetav_1^{\rm test}), \ldots, \model(\thetav_{l}^{\rm test})$ for the emulator based on combined angular and energy distributions with $r^2=0.99$: (a) standardized errors and (b) relative errors. Blue line shows the density of a standard normal random variable.}
        \label{fig:diagnostics_combined}
\end{figure}

In order to test the prediction quality of the trained emulator, we use $m$ parameter sets, which correspond to the $25 \%$ of the parameter sets, namely $\thetav_1^{\rm test}, \ldots, \thetav_m^{\rm test}$, and their corresponding CDCC calculations as test data. We evaluate the performance by computing the test $r^2$ value, standardized error, and the relative error since they are standard ways to evaluate the emulator quality (see App.~\ref{appendix:diag} for the definition of the metrics). The diagnostics plots are illustrated in Fig.~\ref{fig:diagnostics_combined} for fitting the emulator based on the combined angular and energy distributions. For the sake of brevity, we provide the diagnostics plots for the remaining emulators in App.~\ref{appendix:diag}. For all the emulators, the test $r^2$ value is very close to one, and both relative and standardized errors are centered around zero.
          
\section{Results and Discussion}
\label{sec:results}

\begin{figure}[t]
    \begin{center}
        \includegraphics[width=0.4\textwidth]{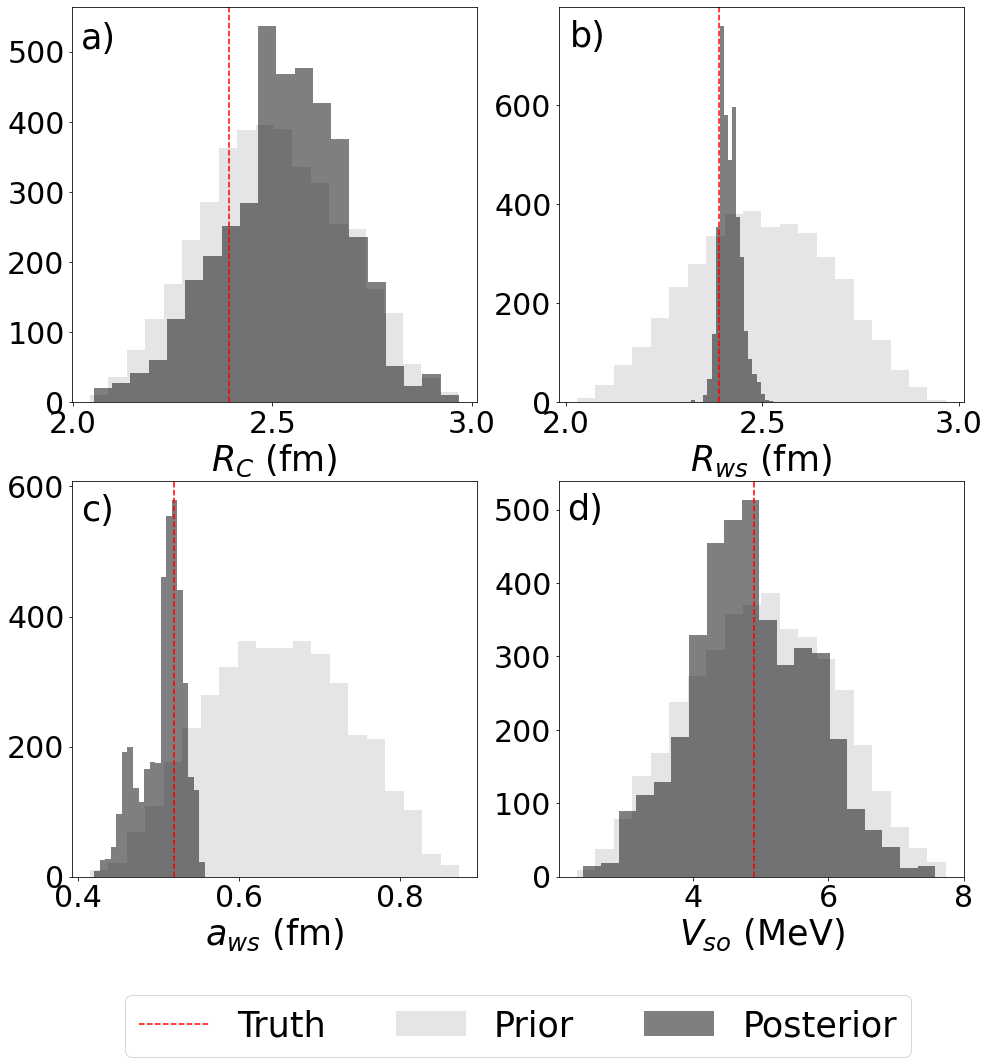}
    \end{center}
    \caption{Univariate marginal estimates of the posterior distribution for the 4-dimensional parameter vector of the breakup reactions for the emulator based on both angular and energy distributions:  prior distributions (light grey) and posterior distributions (dark grey) for the Coulomb radius parameter (a), the Woods-Saxon radius (b), the Woods-Saxon diffuseness (c) and the spin-orbit depth (d). The red dashed line corresponds to the true parameter values.}
    \label{fig-post-combined}
\end{figure}
With trained GP emulators of the computationally expensive CDCC reaction model in hand, we now discuss the posterior distributions obtained for the parameters of the effective interaction between the halo nucleon and the core, as well as the resulting uncertainties on the breakup cross sections.

As mentioned in Sec.~\ref{sec:model}, mock data is generated with the parameters of \cite{capel2007}. Based on the typical statistical errors in breakup experiments of halo nuclei, we include a standard deviation of $10 \%$ on the mock data. Bayesian analysis requires the prior distributions for the core-valence interaction parameters. We use a beta distribution with the shape parameters $\alpha = \beta = 3.5$ as prior for each core-valence parameter $\thetav=(\onetheta_1,\ldots,\onetheta_4)$, meaning
\[
    p(\thetav) = \prod_{i=1}^4 \frac{\Gamma(\alpha + \beta) \bigg(\frac{\onetheta_i - \underline{\onetheta_i}}{\overline{\onetheta_i} - \underline{\onetheta_i}}\bigg)^{\alpha -1}\bigg(1-\frac{\onetheta_i - \underline{\onetheta_i}}{\overline{\onetheta_i} - \underline{\onetheta_i}}\bigg)^{\beta-1}}{\Gamma(\alpha)\Gamma(\beta)},
\]
where $\underline{\onetheta_i}$ and $\overline{\onetheta_i}$ denote, respectively, the lower and upper bounds of the range of parameter $\onetheta_i$ given in Table~\ref{tab:parameters}. The beta distribution is a widely used prior since it is a bounded continuous distribution with a decreasing density at the edges of the parameter range. We implement the parallel-tempering ensemble MCMC algorithm \cite[Chapter 10]{liu2008monte} through the \texttt{surmise} package in order to generate a set of samples from the posterior distribution.

The posteriors obtained by combining both angular and energy distributions are illustrated in Fig.~\ref{fig-post-combined}. For the sake of brevity, the results obtained for angular and energy distributions as separate observables are provided in App.~\ref{appendix:cal}. As compared to the prior distributions represented by the light grey bins, the posterior distributions for $R_{ws}$ and $a_{ws}$ are tightly concentrated around the true parameter values (represented by the red line); this indicates that $R_{ws}$ and $a_{ws}$ are well constrained by the breakup cross section distributions. On the other hand, the posterior distributions of $R_C$ and $V_{so}$ are not very well constrained by this breakup data, implying that those parameters are not very influential in predicting both angular and energy distributions. We did not find significant differences in the results obtained using the other two emulators, which suggests that the information on $V_{cv}$ contained in the breakup energy distribution is similar to that contained in the angular distribution. 

\begin{figure}[t]
    \begin{center}
        \includegraphics[width=0.48\textwidth]{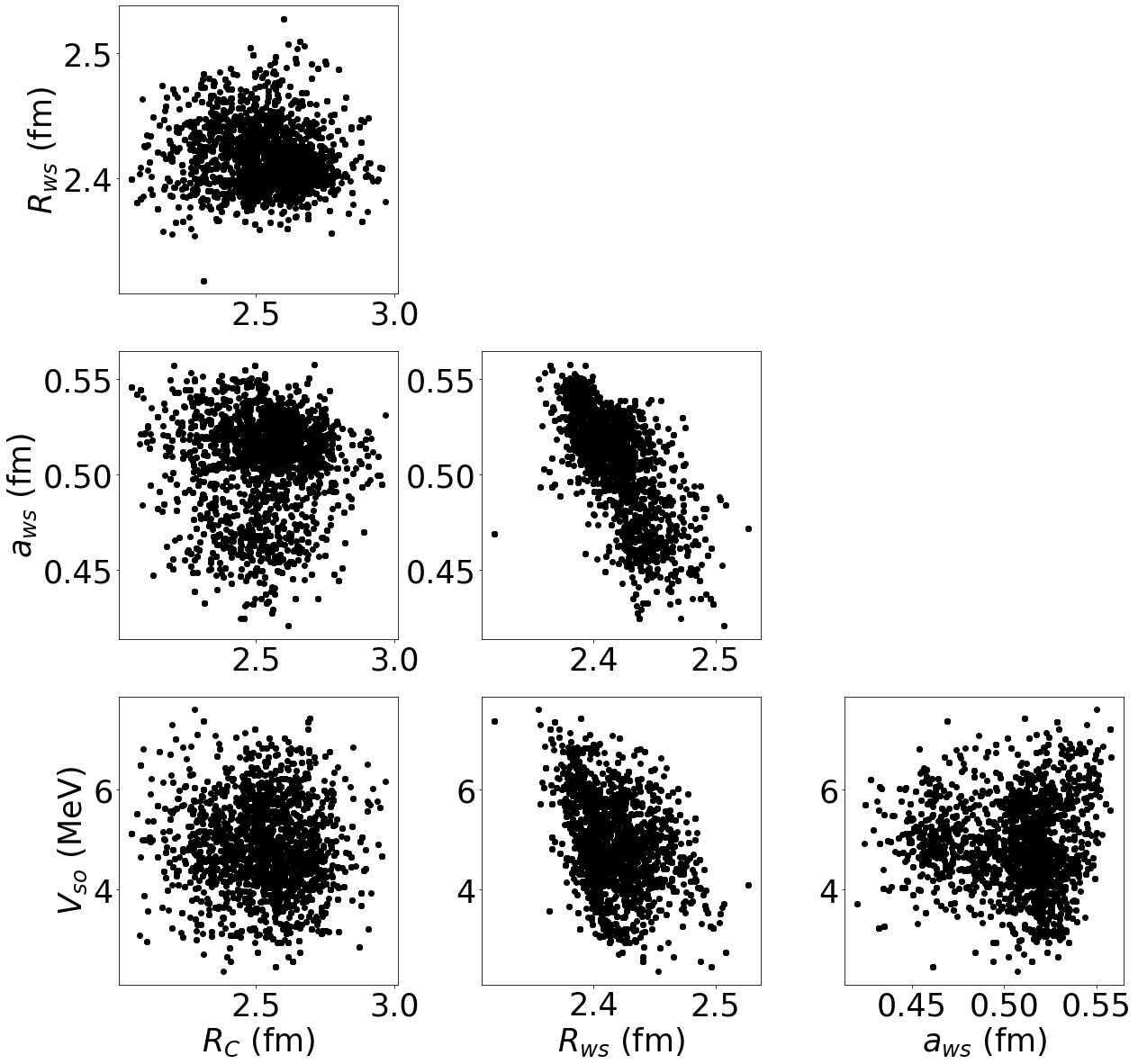}
    \end{center}
    \caption{Pair plots for the posterior samples of four parameters obtained using MCMC with angular and energy distribution data.}
    \label{pairplot_combined}
\end{figure}

Moreover, for the emulator based on the combined angular and energy distributions, we obtain the pair plots using posterior samples of four parameters as in Fig.~\ref{pairplot_combined}. The pair plots indicate a negative correlation between the two most influential parameters $R_{ws}$ and $a_{ws}$.

\begin{figure}[t]
     \begin{subfigure}{0.9\columnwidth}
         \includegraphics[width=1\textwidth]{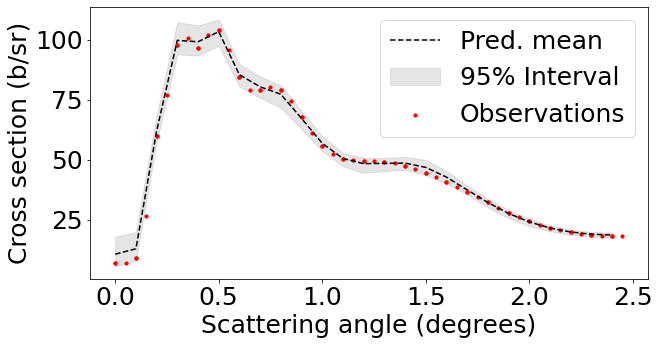}
         \caption{Angular distribution.}
     \end{subfigure}
     \begin{subfigure}{0.9\columnwidth}
         \includegraphics[width=1\textwidth]{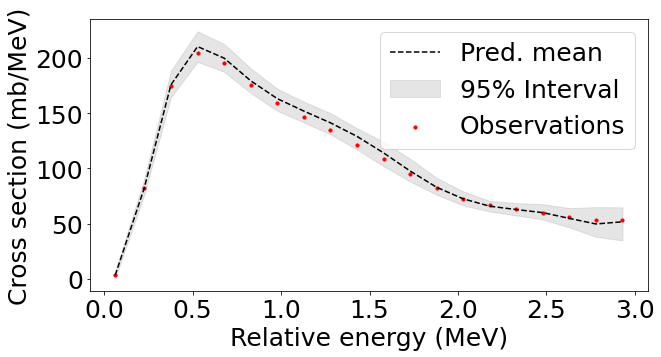}
         \caption{Energy distribution.}
     \end{subfigure}
     \caption{Cross section for $^8$B$ + ^{208}$Pb$ \rightarrow ^7$Be$ + p + ^{208}$Pb at $80$ MeVA (a) angular distribution and (b) relative $^7$Be$+p$ energy distribution: prediction mean (dashed black line) and the 95\% credible interval (shaded gray area) obtained from the Bayesian analysis, compared with mock data. Results correspond to the emulator which uses combined angular and energy distribution data.}
    \label{pred_interval_combined}
\end{figure}
One of the main advantages of using Bayesian techniques is for a rigorous propagation of the quantified uncertainties in the parameters to the predictions. That is, once parameters are sampled from the posterior distribution, they can be used to produce predictions with quantified uncertainties on the breakup cross section distributions. Figure~\ref{pred_interval_combined} displays the $95\%$ credible intervals (grey band), along with the prediction means (black dashed line) and the mock data (red circles) for (a) the breakup angular and (b) the energy distribution. These intervals do not include the uncertainty on the emulation itself. We find the credible interval is roughly within two standard deviations of the predicted mean (corresponding to a $20$\% uncertainty).  

\begin{figure}[ht]
    \begin{center}
        \includegraphics[width=0.3\textwidth]{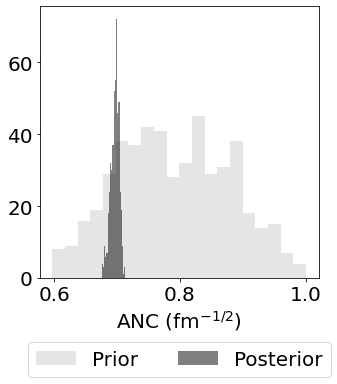}
    \end{center}
\caption{The estimate of the posterior distribution (dark grey bins) of ANC parameters as compared to its prior distribution obtained from training parameters (light grey bins).}
\label{ANCdistribution}
\end{figure}

Finally, from the  posterior distributions shown in Fig.~\ref{fig-post-combined}, we compute the corresponding ANC parameters. Figure~\ref{ANCdistribution} shows the ANC posterior distribution compared to the prior used in the training set. In the model used for the core-valence interaction, the radius and diffuseness are correlated in such a way that the combined breakup data is able to tightly constrain the ANC. This result is consistent with a study on another halo nucleus \cite{capel2006}. From Fig.~\ref{ANCdistribution}, we can extract the value for the $^8$B ANC squared $C^2=0.49 \pm 0.01$ (fm$^{-1}$) which agrees with the value extracted from a different breakup reaction \cite{b8anc}.

\section{Conclusions and Vision} \label{sec:conclusions}

This work consists of the first Bayesian analysis for nuclear breakup reactions.
Here, we focus on the breakup reaction of $^8$B$ + ^{208}$Pb$ \rightarrow ^7$Be$ + p + ^{208}$Pb at $80$ MeVA. The reaction model used is the continuum discretized coupled channel method, inherently a three-body non-perturbative approach to breakup, widely used in the field. Even though there are almost 30  parameters in the model, in this study we consider only the uncertainty from the core-valence effective interaction. The free parameters then are the Coulomb radius $R_C$, the Woods-Saxon radius $R_{ws}$ and diffuseness $a_{ws}$ and the spin-orbit strength $V_{so}$. The depth of the central Woods-Saxon term is fixed by imposing the experimental proton separation energy in the core-valence system. 

In order to perform the Bayesian calibration, we constructed three emulators, one based on the breakup angular distribution, another based on the breakup energy distribution, and a third one based on the combined data set (angular and energy distributions). We used full CDCC calculations to train the emulators and found that approximately 400 training sets were sufficient to obtain an uncertainty on the emulation smaller than the typical error bar on the experimental data ($\approx 10$\%). We used PCA to further reduce the dimensionality of the observable space. We found no significant differences between the predictions obtained with the three emulators. In all three cases, posterior distributions obtained for $R_{ws}$ and $a_{ws}$ were significantly constrained, as opposed to the posterior distributions for $R_C$ and $V_{so}$. Most important, from the Bayesian calibration we obtained $95 \%$ credible intervals for the breakup observables. The resulting uncertainties on the breakup cross sections (both for the angular distribution and the energy distribution) were around $20$\%.

From this study, we also obtained a posterior distribution for the ANC which allows for an extraction of the ANC squared consistent with previous studies. We should emphasize that, although the uncertainty on the extracted ANC is less than $10$\%, this value only includes the uncertainty arising from the core-valence interaction in the three-body model. A larger study, including the uncertainty from the optical potentials, need to be done before the full parametric uncertainty in the model is accounted for.

One can expect further challenges when augmenting the parameter set to include all parameters in the core-target and valence-target optical potential. Obviously, increasing the number of parameters, the parameter space to be explored becomes much larger and the percentage of the parameter space corresponding to a region of high posterior becomes much smaller. In such cases, the training data sets need to be increased considerably to better explore the parameter space \cite{Loeppky2009}. 
Moreover, the computational cost to fit an accurate emulator and run the MCMC sampler grows when the number of uncertain parameters increases. Considering the amount of computational time for a single run of {\sc frescox} (e.g., 25 hours), the one-shot, emulate-then-calibrate principle can lead to an ineffective calibration procedure. In such a case, a sequential Bayesian inference approach depending on active learning procedures may be a better strategy to perform the uncertainty quantification \cite{Jarvenpa2019, Joseph2019}.

\newpage
\appendix

\section{Visualization of cross sections}
\label{appendix:pca}

\begin{figure}[ht]
    \centering
        \includegraphics[width=0.45\textwidth]{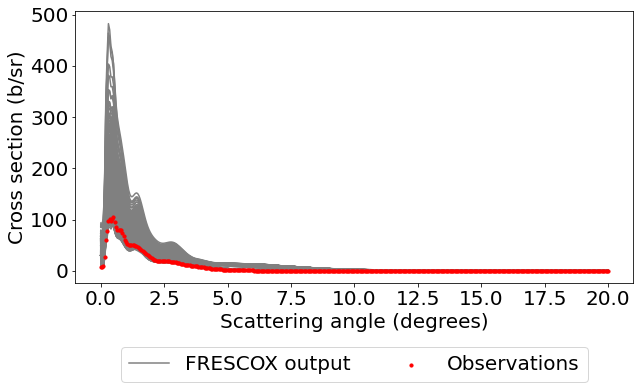}
         \caption{CDCC angular distributions for the breakup of $^8$B on $^{208}$Pb at $80$ MeVA (grey lines) compared to the CDCC results corresponding to the mock data obtained with $R_C=R_{ws}=2.391$ fm, $a_{ws}=0.52$ fm, and $V_{so}=4.898$ MeV (red circles).}
         \label{outcome_angular}
\end{figure}

\begin{figure}[ht]
     \begin{subfigure}{1\columnwidth}
        \includegraphics[width=1\textwidth]{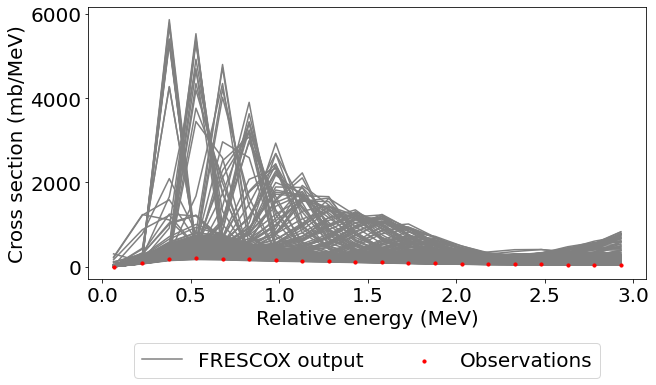}
         \caption{Simulation outputs before filtering.}
         \label{nofilter}
     \end{subfigure}
     \begin{subfigure}{1\columnwidth}
         \includegraphics[width=1\textwidth]{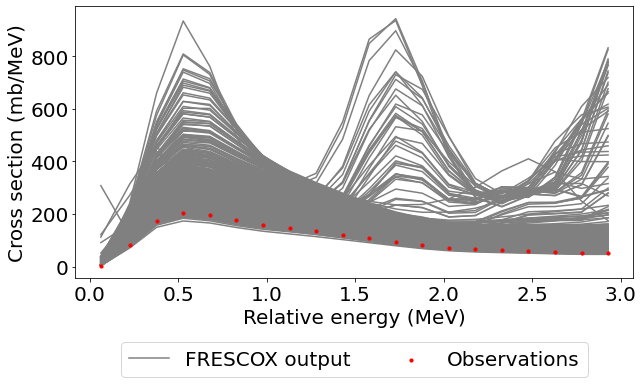}
         \caption{Simulation outputs after filtering.}
         \label{filter}
     \end{subfigure}
     \caption{Same as Fig.~\ref{outcome_angular}  for the energy distributions: (a) before filtering and (b) after filtering.}
     \label{outcome_energy}
\end{figure}

In this appendix, for completeness, we present the information regarding the CDCC simulation cross sections used to train the emulators. Using the parameter values sampled from the LHS as discussed in Sec.~ \ref{sec:emulators} (illustrated in Fig.~\ref{fig-sample}), we obtain a set of angular distributions (Fig.~\ref{outcome_angular}) and energy distributions (Fig.~\ref{outcome_energy}). Figure~\ref{nofilter} contains all the outputs and Fig.~\ref{filter} contains the set after filtering out the simulation outputs that drastically differ from the mock data to increase the predictive accuracy of the emulators. It was the set of cross sections in Fig.~\ref{filter} that was ultimately used in the emulator based on the energy distributions.
 
\section{Gaussian-process based emulator}
\label{appendix:emulator}

In this appendix we explain how the emulator is constructed. The first step involves the reduction of dimensionality through PCA.
PCA represents the data matrix $\Xi$ in $q$-dimensional space as $\Wv = \Bv^\top \Xi$, where $\Wv$ is the $q \times n$ score matrix of principal components in the latent space and the $d \times q$ matrix $\Bv = [\bv_1, \ldots, \bv_q]$ stores the orthogonal basis vectors $\{\bv_1, \ldots, \bv_q\}$. Once we obtain $\Wv$ via PCA, we fit a GP for each of $q$ latent outputs (i.e., each row of $\Wv$) as follows. To construct an emulator, the simulation output is modeled using the basis representation of \cite{Higdon2008}:
\begin{equation}
    \model(\thetav) = \sum_{i=1}^q \bv_i w_i(\thetav) + \pcaepsilonv,
    \label{eq:basis_rep}
\end{equation}
where $w_i(\thetav)$ are weights modeled with independent, univariate GPs and $\pcaepsilonv$ is a $d$-dimensional term that contains the residual between the model and the span of the orthogonal basis vectors $\{\bv_1, \ldots, \bv_q\}$.

To construct a GP for a function mapping $w_i(\thetav)$, following \cite{Rasmussen2005}, we start by assuming a GP prior on the function $w_i(\thetav)$ as provided in Eq.~\ref{eq:gp_prior}. The covariance function is the key ingredient of a GP as it determines the similarity of outputs for different parameters. Following \cite{gramacy2012cases}, the covariance between two outputs is denoted as
\begin{equation}
    C_i(\thetav, \thetav') = \tau_{i}^{2} K_i(\thetav, \thetav') = \tau_{i}^{2} \left[R(\thetav, \thetav'; \rhov_{i}) + \upsilon_{i}\delta_{\thetav, \thetav'} \right],
    \label{eq:covariance}
\end{equation}
where $K_i(\thetav, \thetav') \coloneqq R(\thetav, \thetav'; \rhov_{i}) + \upsilon_{i}\delta_{\thetav, \thetav'}$ and $\tau_{i}^{2}$ represents a scaling parameter, $R(\thetav, \thetav'; \rhov_{i})$ is a correlation function, $\upsilon_{i}$ is a nugget parameter, and $\delta_{\thetav, \thetav'}$ is a Kronecker delta function which is 1 if $\thetav=\thetav'$. Here, $\tau_{i}$, $\upsilon_{i}$, and $\rhov_{i}$ are unknown hyperparameters, and estimated as described in App.~\ref{appendix:hyper}. For the correlation function $R(\cdot)$, in this work we use the separable version of the Mat\'ern correlation function with smoothness parameter 1.5 \cite{Rasmussen2005} such that
\begin{equation}
    \begin{aligned}
        R(\thetav, \thetav'; \rhov_{i}) = & \bigg[\prod_{j=1}^p (1 + |(\thetav_j - \thetav_j')\exp(\rhov_{i,j})|)\bigg] \times \\ & \exp\bigg(-\sum_{j=1}^p |(\thetav_j - \thetav_j')\exp(\rhov_{i,j})|\bigg).
    \end{aligned}
\end{equation} 
The hyperparameter vector $\rhov_i = (\rhov_{i,1}, \ldots, \rhov_{i,p})$ controls the correlation strength along each dimension of the latent output $w_i(\cdot)$. The Mat\'ern correlation function is a popular choice for GPs since it has differentiable sample paths \cite{handcock1993bayesian}.

Recall that the $n$-vector $\wv_i$ stores $\wv_i = (w_i(\thetav_1^{\rm tr}), \ldots, w_i(\thetav_{n}^{\rm tr}))$ for $i = 1, \ldots, q$. We can also write the prior for $\wv_i$ as $\wv_i \sim \text{N}(\gammav_{i}, \Cv_{i})$. We assume a constant mean for each GP such that $\gammav_{i} = \gamma_{i} \bone_n$, and we estimate the unknown hyperparameter $\gamma_{i}$ in addition to $\tau_{i}$, $\upsilon_{i}$, and $\rhov_{i}$ as described in the next section.

\section{Hyperparameter estimation}
\label{appendix:hyper}

There are many ways to infer the parameters $\gamma_i$, $\tau_{i}$, $\upsilon_{i}$, and $\rhov_{i}$ given training data $\{\thetav_1^{\rm tr}, \ldots, \thetav_{n}^{\rm tr}\}$ and $\{\model(\thetav_1^{\rm tr}), \ldots,$ $\model(\thetav_{n}^{\rm tr})\}$. One common way is to maximize the resulting likelihood due to $\wv_i \sim \text{N}(\gammav_{i}, \Cv_i)$. In this study, we use the maximum likelihood estimation as it is computationally more efficient than their Bayesian counterparts. The multivariate normal distribution results in the likelihood
\begin{equation}
    \frac{|\Kv_{i}|^{-1/2}}{(2 \pi \tau_{i}^{2})^{n/2}}  \exp{\bigg\{-\frac{1}{2\tau_{i}^{2}} (\wv_i - \gammav_{i})^\top \Kv_{i}^{-1} (\wv_i - \gammav_{i}) \bigg\}},
    \label{likelihoodhyper}
\end{equation}
with the $n \times n$ matrix $\Kv_{i}$ obtained from applying $K_i(\cdot, \cdot)$ in Eq.~\ref{eq:covariance} to each pair of the parameter settings $\thetav_1^{\rm tr}, \ldots, \thetav_{n}^{\rm tr}$. Taking the log of Eq.~\ref{likelihoodhyper} and applying $\gammav_{i} = \gamma_{i} \bone_n$ yields the log-likelihood
\begin{equation}
    \begin{aligned}
     = & - \frac{n}{2} \log(2\pi) - \frac{n}{2} \log(\tau_{i}^{2}) - \frac{1}{2} \log(|\Kv_{i}|)  \\ & - \frac{1}{2\tau_{i}^{2}} (\wv_i - \gamma_{i} \bone_n)^\top \Kv_{i}^{-1} (\wv_i - \gamma_{i} \bone_n).
    \label{eq:hyper_log_1}
    \end{aligned}
\end{equation}

The parameters $\gamma_{i}$ and $\tau_{i}^{2}$ are obtained by maximizing the log-likelihood, yielding
\begin{equation}
    \begin{aligned}
        \hat{\gamma}_{i} &= \frac{\bone_n^\top \Kv_{i}^{-1} \wv_i }{\bone_n^\top \Kv_{i}^{-1} \bone_n} \text{ and } \\ \hat{\tau}_{i}^{2} &= \frac{1}{n} (\wv_i - \hat{\gamma}_{i} \bone_n)^\top \Kv_{i}^{-1} (\wv_i - \hat{\gamma}_{i} \bone_n).
    \end{aligned}
    \label{eq:hyper_log_2}
\end{equation}

Plugging the estimates $\hat{\gamma}_{i}$ and $\hat{\tau}_{i}^{2}$ in Eq.~\ref{eq:hyper_log_2} into the log-likelihood in Eq.~\ref{eq:hyper_log_1} gives a revised log-likelihood of
\begin{equation}
    \begin{aligned}
        l^\star(\upsilon_{i}, \rhov_{i}) =  - \frac{n}{2} \log(\hat{\tau}_{i}^{2}) - \frac{1}{2} \log(|\Kv_{i}|),
    \end{aligned}    
    \label{eq:revised_log_like}
\end{equation}
where the additive constant has been dropped.
We then employ \texttt{scipy}'s implementation of L-BFGS-B \cite{Nocedal2006} to estimate $\upsilon_{i}$ and $\rhov_{i}$ by maximizing $l^\star$. With the estimates for parameters $\gamma_i$, $\tau_{i}$, $\upsilon_{i}$, and $\rhov_{i}$ in hand, we can use $q$ independent GPs to make predictions on the simulator’s output for unseen points as described in Sec.~\ref{sec:predictions}.

\section{Diagnostics plots}
\label{appendix:diag}

\begin{figure}[h]
     \begin{subfigure}{0.48\columnwidth}
         \includegraphics[width=1\textwidth]{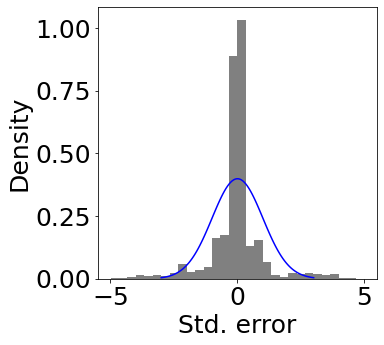}
         \caption{Standardized errors.}
     \end{subfigure}
     \begin{subfigure}{0.48\columnwidth}
         \centering
         \includegraphics[width=1\textwidth]{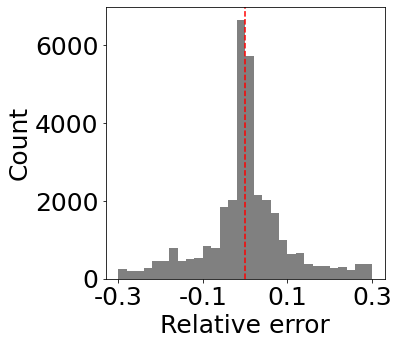}
         \caption{Relative errors.}
     \end{subfigure}
        \caption{Diagnostic plots to check the quality of the emulator using test parameters $\thetav_1^{\rm test}, \ldots, \thetav_{l}^{\rm test}$ and their corresponding simulation outputs $\model(\thetav_1^{\rm test}), \ldots, \model(\thetav_{l}^{\rm test})$ for the emulator based on angular distribution with $r^2=1$: (a) standardized errors and (b) relative errors. Blue line shows the density of a standard normal random variable.}
        \label{fig:diagnostics_angular}
\end{figure}

\begin{figure}
     \begin{subfigure}{0.48\columnwidth}
         \includegraphics[width=1\textwidth]{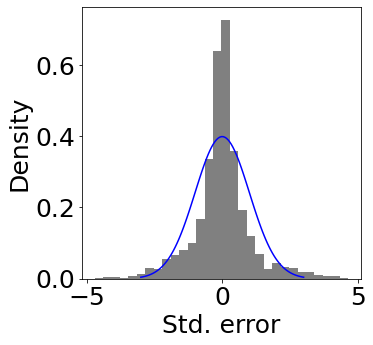}
         \caption{Standardized errors.}
     \end{subfigure}
     \begin{subfigure}{0.48\columnwidth}
         \centering
         \includegraphics[width=1\textwidth]{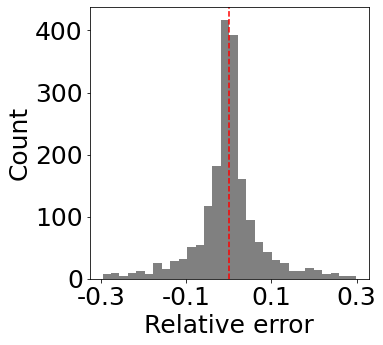}
         \caption{Relative errors.}
     \end{subfigure}
        \caption{Diagnostic plots to check the quality of the emulator using test parameters $\thetav_1^{\rm test}, \ldots, \thetav_{l}^{\rm test}$ and their corresponding simulation outputs $\model(\thetav_1^{\rm test}), \ldots, \model(\thetav_{l}^{\rm test})$ for the emulator based on energy distribution with $r^2=0.96$: (a) standardized errors and (b) relative errors. Blue line shows the density of a standard normal random variable.}
        \label{fig:diagnostics_energy}
\end{figure}

In this study, we fit three different emulators using breakup angular $\angular$, energy $\energy$, and both angular $\angular$ and energy $\energy$ distributions as observables. Once we fit the emulators, we calibrate the computer model via MCMC using each emulator as an input to the calibrator. Before we integrate the final emulators into the calibration process, we make sure that the cross sections are well emulated. To do that, we randomly separate out $25\%$ of data to use as test data, and fit each of the emulators with the remaining data. We then evaluate the performance of the emulator by computing the test $r^2$ value, standardized error, and the relative error via
\begin{equation}
    \begin{aligned}
        r^2 &= 1 - \frac{\sum\limits_{i = 1}^{m}\sum\limits_{j=1}^{d} (\model_j(\thetav^{\rm test}_i) - \muv_j^{\rm GP}(\thetav^{\rm test}_i))^2}{\sum\limits_{i = 1}^{l}\sum\limits_{j=1}^{d}(\model_j(\thetav^{\rm test}_i) - \bar{\model}_j)^2},  \\
        & \frac{\model_j(\thetav^{\rm test}_i) - \muv_j^{\rm GP}(\thetav^{\rm test}_i)}{\sqrt{{\rm diag}_j(\Sigmav^{\rm GP}(\thetav^{\rm test}_i))}}, \text{and }
        1 - \frac{\muv_j^{\rm GP}(\thetav^{\rm test}_i)}{\model_j(\thetav^{\rm test}_i)},
    \end{aligned}
\end{equation}
respectively, for $i = 1, \ldots, m$ and $j = 1, \ldots, d$. Here, ${\rm diag}_j(\Sigmav^{\rm GP}(\thetav^{\rm test}_i))$ represents the diagonal elements of the emulator covariance matrix, namely, the predictive variances, and the index $j$ is used to denote the $j$th element of the corresponding vectors. In addition, $\bar{\model}_j$ is the average simulation output for the $j$th input dimension.
The main text contained the diagnostic results for the emulator based on the combined data (angular and energy distributions). In this appendix we present the diagnostic results for the emulator based on the angular distributions alone (Fig.~\ref{fig:diagnostics_angular}) and for the emulator based on the energy distributions alone (Fig.~\ref{fig:diagnostics_energy}). Our results show that all emulators have similar performance and are able to reproduce the CDCC simulations within $10$\%.

\section{Constraining breakup angular and energy distributions}
\label{appendix:cal}

\begin{figure}
    \begin{center}
        \includegraphics[width=0.4\textwidth]{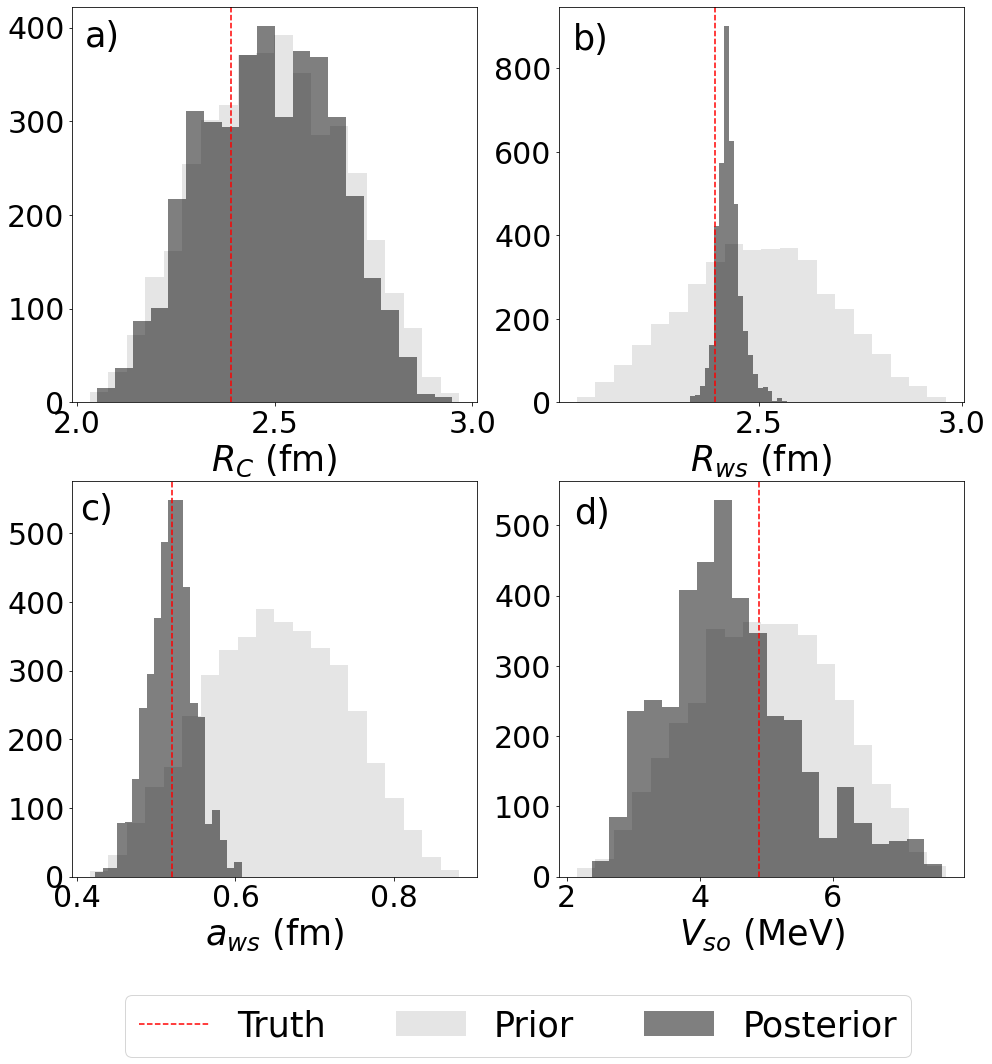}
    \end{center}
    \caption{Univariate marginal estimates of the posterior distribution for the 4-dimensional parameter vector of the breakup reactions for the emulator based on angular distributions: prior distributions (light grey) and posterior distributions (dark grey) for the Coulomb radius parameter (a), the Woods-Saxon radius (b), the Woods-Saxon diffuseness (c) and the spin-orbit depth (d). The red dashed line corresponds to the true parameter values.}
    \label{fig-post-angular}
\end{figure}
\begin{figure}[b]
    \begin{center}
        \includegraphics[width=0.4\textwidth]{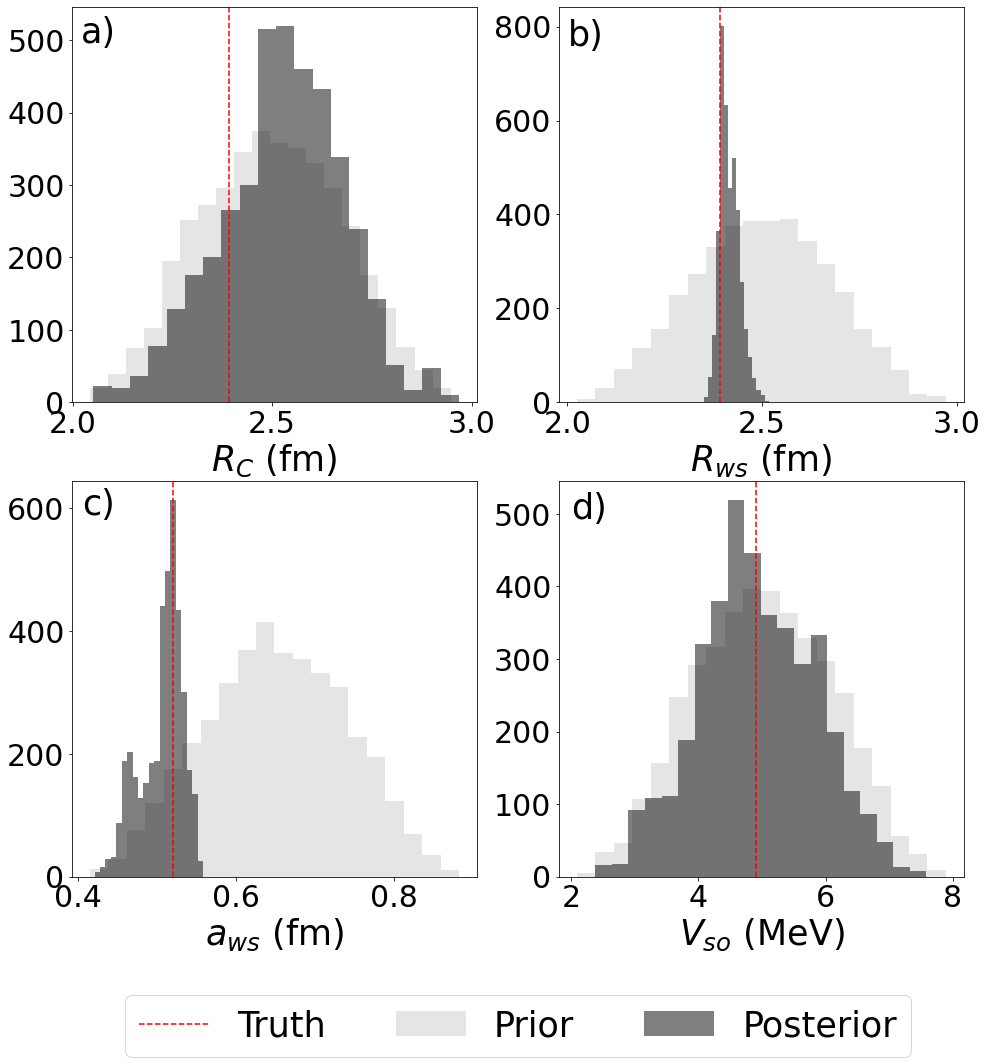}
    \end{center}
    \caption{Univariate marginal estimates of the posterior distribution for the 4-dimensional parameter vector of the breakup reactions for the emulator based on energy distributions:  prior distributions (light grey) and posterior distributions (dark grey) for the Coulomb radius parameter (a), the Woods-Saxon radius (b), the Woods-Saxon diffuseness (c) and the spin-orbit depth (d). The red dashed line corresponds to the true parameter values.}
    \label{fig-post-energy}
\end{figure}
In this appendix, we provide the calibration results for the emulators that are trained with either $\angular$ or $\energy$ distributions. 
We first consider the posterior distribution for the parameters. Figure~\ref{fig-post-angular} corresponds to the emulator based on angular distributions while Fig.~\ref{fig-post-energy} corresponds to the results obtained with the emulator trained with energy distributions. As discussed in Sec.~\ref{sec:results}, and for both these emulators, only $R_{ws}$ and $a_{ws}$ are well constrained and, as expected, their posterior distributions are centered around the parameter values used to produce the mock data.

\begin{figure}
    \begin{center}
        \includegraphics[width=0.45\textwidth]{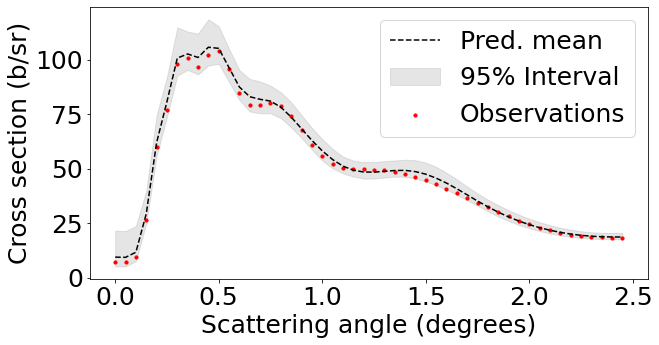}
    \end{center}
\caption{Cross section angular distributions for $^8$B$ + ^{208}$Pb$ \rightarrow ^7$Be$ + p + ^{208}$Pb at $80$ MeVA using the emulator trained on angular distributions alone: prediction mean (dashed black line) and the 95\% credible interval (shaded gray area) obtained from the Bayesian analysis, compared with mock data.}
\label{fig-xsangle}
\end{figure}
\begin{figure}
    \begin{center}
        \includegraphics[width=0.45\textwidth]{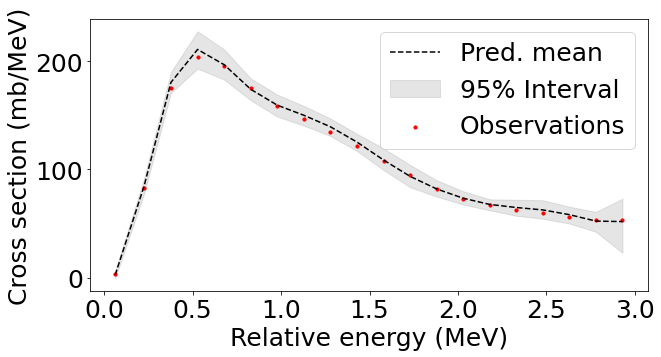}
    \end{center}
\caption{Cross section energy distributions for $^8$B$ + ^{208}$Pb$ \rightarrow ^7$Be$ + p + ^{208}$Pb at $80$ MeVA using the emulator trained on energy distributions alone: prediction mean (dashed black line) and the 95\% credible interval (shaded gray area) obtained from the Bayesian analysis, compared with mock data.}
\label{fig-xsenergy}
\end{figure}
We next consider the observables generated with the two emulators. The emulator trained on angular distributions is only capable of predicted angular distributions while the emulator trained on energy distributions is only able to predict energy distributions.
Figure~\ref{fig-xsangle} shows the confidence intervals obtained for the cross sections with the emulator trained on angular distributions and Fig.~\ref{fig-xsenergy} shows the confidence intervals obtained for the cross section with the emulator trained on energy distributions.
These results are quantitatively similar to the results presented in Sec.~\ref{sec:results} for the emulator based on the combined set of data (both angular and energy distributions). 

\begin{acknowledgments}

This material is based upon work supported by the National Science Foundation CSSI program under award number OAC-2004601 (BAND Collaboration). We gratefully acknowledge the computing resources provided on Bebop, a high-performance computing cluster operated by the Laboratory Computing Resource Center at Argonne National Laboratory.
This work was supported in part by the U.S.\ Department of Energy, Office of Science, Nuclear Physics program (under grant DE-SC0021422) and Advanced Scientific Computing Research NUCLEI SciDAC (under grant DE-AC02-06CH11357), 
and by the National Science Foundation (under grant PHY-1811815).

\end{acknowledgments}

\newpage
\section*{Bibliography}
\bibliography{main}

\end{document}